\documentclass[preprint,showpacs,pra]{revtex4}

\usepackage{graphicx}
\usepackage{dcolumn}
\usepackage{bm}
\usepackage{amsfonts}

\newcommand{\mycomm}[1]{}

\newcommand{\pr}{Phys. Rev. }
\newcommand{\jpb}{J. Phys. B }

\newcommand{\UQ}{ARC Centre of Excellence for Quantum-Atom Optics, School of Physical Sciences, University of Queensland, Brisbane, Qld 4072, Australia.}

\newcommand{\etal}{{\em et al.}}

\begin{document}

\title{Tripartite entanglement from interlinked $\chi^{(2)}$ parametric interactions}

\author{M.~K. Olsen} 

\affiliation{\UQ}

\author{A.~S. Bradley}

\affiliation{\UQ}

\begin{abstract}
We examine the tripartite entanglement properties of an optical system using interlinked $\chi^{(2)}$ interactions, recently studied experimentally in terms of its phase-matching properties by Bondani \etal\, [M. Bondani, A. Allevi, E. Gevinti, A. Agliati, and A. Andreoni, arXiv:quant-ph/0604002.]. We show that the system does produce output modes which are genuinely tripartite entangled and that detection of this entanglement depends crucially on the correlation functions which are measured, with a three-mode Einstein-Podolsky-Rosen inequality being the most sensitive.
\end{abstract}

\date{\today}

\maketitle

\section{Introduction}
\label{sec:intro}

Entanglement is a property which is central to quantum mechanics and helps to distinguish it from classical mechanics. A vast amount of work has been undertaken on discrete variable entanglement, with somewhat less having been performed on the continuous variable case. 
It is the latter which interests us in this work, particularly as regards the entanglement of three optical modes. We will focus on an experimentally realised system which links two $\chi^{(2)}$ interactions in a combined downconversion and sum frequency generation process~\cite{Bondani}, and examine its utility for the production of states which exhibit full tripartite entanglement.
As far as we are aware, full tripartite entanglement has only been unambiguously demonstrated by mixing squeezed vacua with linear optical elements~\cite{Jing,aoki}, although other methods which create the entanglement using an actual nonlinear interaction are under investigation, using both cascaded and concurrent $\chi^{(2)}$ processes~\cite{Guo,Ferraro,Nosso,nl2}.

The definition of tripartite entanglement for three-mode systems is a little more subtle than that for bipartite entanglement, with different classes of entanglement having been defined, depending on how the system density matrix may be partitioned~\cite{Giedke}. The classifications range from fully inseparable, which means that the density matrix is not separable for any grouping of the modes, to fully separable, where the three modes are not entangled in any way. For the fully inseparable case, van Loock and Furusawa~\cite{vanLoock2003}, who call this genuine tripartite entanglement, have derived inequalities which are easily applicable to continuous variable processes. More recently, Olsen \etal~\cite{nl3} have defined three-mode Einstein-Podolsky-Rosen (EPR)~\cite{EPR} type criteria, which also provide sufficient, but not necessary, conditions for the demonstration of genuine tripartite entanglement. 
In this article we will begin by reviewing the definitions of these entanglement criteria and then apply them to the outputs of the Bondani scheme to quantify entanglement correlations which may in principle be measured experimentally.

\section{Criteria for tripartite entanglement}
\label{sec:criteria}

We begin by giving the optical quadrature definitions we will use in our analysis, as the exact form of the inequalities will depend on these. For three modes described by the bosonic annihilation operators $\hat{a}_{j}$, where $j=1,2,3$, we define 
quadrature operators for each mode as
\begin{equation}
\hat{X}_j = \hat{a}_j+\hat{a}_j^\dag,\:\:\:
\hat{Y}_j = -i(\hat{a}_j-\hat{a}_j^\dag),
\label{eq:quaddefs}
\end{equation}
so that the Heisenberg uncertainty principle requires $V(\hat{X}_{j})V(\hat{Y}_{j})\geq 1$. 

\subsection{Three-mode Einstein-Podolsky-Rosen correlations}
\label{subsec:EPR3}

The EPR argument was introduced in 1935 in an attempt to show that quantum mechanics could not be both complete and consistent with local realism~\cite{EPR}. Schr\"odinger replied that same year by introducing the concept of entangled states which were not compatible with classical notions such as local realism~\cite{gato}. In 1989 Reid~\cite{mdr1}, and Reid and Drummond~\cite{mdr1b} proposed a physical test of the EPR paradox using optical quadrature amplitudes, which are mathematically identical to the position and momentum originally considered by EPR. Reid later expanded on this work, demonstrating that the satisfaction of the 1989 two-mode EPR criterion always implies bipartite quantum entanglement~\cite{mdr2}. Tan made a similar demonstration in the context of teleportation, considering the outputs from a nondegenerate optical parameteric amplifier (OPA) mixed on a beamsplitter~\cite{Sze}. In this article we use an extension of Reid's original approach to the case of tripartite correlations, where quadratures of three different optical modes are involved. This extension was developed and formally proven to demonstrate the presence of tripartite entanglement by Olsen, Bradley and Reid in Ref.~\cite{nl3}, so that we shall call these the OBR criteria. 

There are two ways to consider the experimentally accessible form of the OBR criteria, depending on whether we use information from two quadratures to infer properties of the other, or information from one to infer combined properties of the other two. 
In the first case we make a linear estimate of the quadrature $\hat{X}_i$ from the properties of the
combined mode $j+k$, using parameters which can be optimised, both experimentally and theoretically~\cite{mdr1,Ou}. It has been shown~\cite{nl3,mdr1b} that minimising the root mean square error in this estimate leads to an optimal inferred variance,
\begin{eqnarray}
V^{inf}(\hat{X}_{i}) &=& V(\hat{X}_{i})-\frac{\left[V(\hat{X}_{i},\hat{X}_{j}\pm \hat{X}_{k})\right]^{2}}{V(\hat{X}_{j}\pm \hat{X}_{k})},
\end{eqnarray}
where $V(\hat{A},\hat{B})=\langle \hat{A}\hat{B}\rangle-\langle \hat{A}\rangle\langle \hat{B}\rangle$.
We follow the same procedure for the $\hat{Y}$ quadratures to give expressions which may be obtained by swapping each $\hat{X}$ for a $\hat{Y}$ in the above to give the optimal inferred estimate
\begin{eqnarray}
V^{inf}(\hat{Y}_{i}) &=& V(\hat{Y}_{i})-\frac{\left[V(\hat{Y}_{i},\hat{Y}_{j}\pm \hat{Y}_{k})\right]^{2}}{V(\hat{Y}_{j}\pm \hat{Y}_{k})}.
\label{eq:EPR2}
\end{eqnarray}
A demonstration of the EPR paradox can be claimed whenever theory predicts
\begin{equation}
V^{inf}(\hat{X}_{i})V^{inf}(\hat{Y}_{i}) < 1.
\label{eq:infer21}
\end{equation}
As was proven~\cite{nl3}, this demonstration for the $3$ possible values of $i$ is then sufficient to establish tripartite entanglement, without any assumptions having been made about whether the states involved are Gaussian or not.

Following the same logic, if we use the properties of mode $i$ to infer properties of the combined mode $j+k$, we find that
there is a demonstration of the other three mode form of the EPR paradox whenever
\begin{equation}
V^{inf}(\hat{X}_{j}\pm \hat{X}_{k})V^{inf}(\hat{Y}_{j}\pm \hat{Y}_{k})< 4,
\label{eq:infer12}
\end{equation}
where, for example,
\begin{equation}
V^{inf}(\hat{X}_{j}\pm \hat{X}_{k})=V(\hat{X}_{j}\pm \hat{X}_{k})-\frac{\left[V(\hat{X}_{i},\hat{X}_{j}\pm \hat{X}_{k})\right]^{2}}{V(\hat{X}_{i})}.
\end{equation}
As above, this demonstration for the $3$ possible combinations also serves to establish complete inseparability of the density matrix.

\subsection{The van Loock-Furusawa inequalities}
\label{subsec:vLF}

A set of conditions which are sufficient to demonstrate tripartite entanglement for any quantum state have been derived by van Loock and Furusawa~\cite{vanLoock2003}. Using our quadrature definitions, the van Loock-Furusawa conditions give a set of inequalities, which we shall refer to as the VLF inequalities,
\begin{eqnarray}
V_{12} &=& V(\hat{X}_1-\hat{X}_2) + V(\hat{Y}_1+\hat{Y}_2+g_{3}\hat{Y}_3) \geq 4,\nonumber\\
V_{13} &=& V(\hat{X}_1-\hat{X}_3) + V(\hat{Y}_1+g_{2}\hat{Y}_2+\hat{Y}_3) \geq 4,\nonumber\\
V_{23} &=& V(\hat{X}_2-\hat{X}_3) + V(g_{1}\hat{Y}_1+\hat{Y}_2+\hat{Y}_3) \geq 4,
\label{eq:tripart}
\end{eqnarray}
where $V(A)\equiv\langle A^2\rangle-\langle A\rangle^2$ and the $g_{i}$ are arbitrary real numbers.
As shown in reference~\cite{vanLoock2003}, the violation of the first inequality still leaves the possibility that mode $3$ could be separated from modes $1$ and $2$, but this possibility is negated by violation of the second. Therefore, if any two of these inequalities are violated, the system is fully inseparable and genuine tripartite entanglement is guaranteed. We note also that genuine tripartite entanglement may still be possible when none of these inequalities is violated.

We will now investigate optimisation of the VLF criteria, using the freedom allowed in the choice of the $g_{i}$, which are arbitrary real parameters. A simple minimisation of the right-hand sides of Eq.~\ref{eq:tripart} with respect to the $g_{i}$ gives
\begin{eqnarray}
g_{1} &=& \frac{-(\langle\hat{Y}_{1}\hat{Y}_{2}\rangle+\langle\hat{Y}_{1}\hat{Y}_{3}\rangle)}{\langle\hat{Y}_{1}^{2}\rangle},\nonumber\\
g_{2} &=& \frac{-(\langle\hat{Y}_{1}\hat{Y}_{2}\rangle+\langle\hat{Y}_{2}\hat{Y}_{3}\rangle)}{\langle\hat{Y}_{2}^{2}\rangle},\nonumber\\
g_{3} &=& \frac{-(\langle\hat{Y}_{1}\hat{Y}_{3}\rangle+\langle\hat{Y}_{2}\hat{Y}_{3}\rangle)}{\langle\hat{Y}_{3}^{2}\rangle}.
\label{eq:optimise}
\end{eqnarray}
The required variances can now be written as, for example,
\begin{eqnarray}
V(\hat{X}_{1}-\hat{X}_{2}) &=& \langle\hat{X}_{1}^{2}\rangle+\langle\hat{X}_{2}^{2}\rangle-2\langle\hat{X}_{1}\hat{X}_{2}\rangle,\nonumber\\
V(\hat{Y}_{1}+\hat{Y}_{2}+g_{3}\hat{Y}_{3}) &=& \langle\hat{Y}_{1}^{2}\rangle+\langle\hat{Y}_{2}^{2}\rangle+g_{3}^{2}\langle\hat{Y}_{3}^{2}\rangle+2\left[\langle\hat{Y}_{1}\hat{Y}_{2}\rangle+g_{3}
\left(\langle\hat{Y}_{1}\hat{Y}_{3}\rangle+\langle\hat{Y}_{2}\hat{Y}_{3}\rangle\right)\right].
\label{eq:VLFmoments}
\end{eqnarray}
Once this optimisation process has taken place, we find that, for example,
\begin{equation}
V(\hat{Y}_{1}+\hat{Y}_{2}+g_{3}\hat{Y}_{3}) = V(\hat{Y}_{1}+\hat{Y}_{2})-\frac{\left[V(\hat{X}_{3},\hat{Y}_{1}+\hat{Y}_{2})\right]^{2}}{V(\hat{Y}_{3})},
\end{equation}
where we recognise the right hand side as an inferred variance as introduced in Ref.~\cite{nl3} to demonstrate the EPR paradox for three modes, and referred to above. 
The optimised correlations can now be written as
\begin{eqnarray}
V_{12} &=& V(\hat{X}_1-\hat{X}_2) + V^{inf}(\hat{Y}_1+\hat{Y}_2) \geq 4,\nonumber\\
V_{13} &=& V(\hat{X}_1-\hat{X}_3) + V^{inf}(\hat{Y}_1+\hat{Y}_3) \geq 4,\nonumber\\
V_{23} &=& V(\hat{X}_2-\hat{X}_3) + V^{inf}(\hat{Y}_2+\hat{Y}_3) \geq 4.
\label{eq:opttripart}
\end{eqnarray}
We see that the VLF criteria now have the same form as the Duan and Simon criteria for bipartite entanglement~\cite{Duan,Simon}, but with the actual variance $V(\hat{Y}_{j}+\hat{Y}_{k})$ replaced by the inferred variance $V^{inf}(\hat{Y}_{j}+\hat{Y}_{k})$ of Eq.~\ref{eq:infer12}.
We note that the violation of two out of three of the inequalities is sufficient to demonstrate full inseparability. 

\section{System and equations of motion}
\label{sec:hem}

The interaction Hamiltonian used by Bondani \etal~\cite{Bondani} uses an undepleted pumps approximation and we will begin with a more complete form which quantises all the interacting fields. 
The Hamiltonian describes the coupling of five modes of the electromagnetic field in a phase-matched simultaneous sum frequency generation and downversion process  in a manner analogous to the schemes considered by Olsen and Bradley~\cite{nl2}, and has previously been investigated by Ferraro \etal~\cite{Ferraro} and Smithers and Lu~\cite{Smithers}. In Ref.~\cite{nl2}, this five-mode Hamiltonian is written as
\begin{equation}
{\cal H}_{int}=i\hbar\left(\chi_{1}\hat{a}_{4}\hat{a}^{\dag}_{1}\hat{a}_{3}^{\dag}+\chi_{2}\hat{a}_{5}\hat{a}_{2}^{\dag}\hat{a}_{3}\right)+{\mbox h.c.},
\label{eq:parma}
\end{equation}
once we change the indices to agree with Bondani \etal~\cite{Bondani} and set the coupling coefficients to be real. Due to energy conservation, $\omega_{4}=\omega_{1}+\omega_{3}$ and $\omega_{2}=\omega_{3}+\omega_{5}$, and the necessary phae-matching conditions are covered in Ref.~\cite{Bondani}.
This Hamiltonian approximately describes a downconversion process cascaded with a sum-frequency generation process where one of the downconverted modes becomes an auxiliary pump mode for the frequency generation process. It gives a simplified description because it does not include effects such as dispersion within the nonlinear medium, for example. A more accurate method of analysing these types of processes has been given by Raymer \etal~\cite{Raymer}, but the approximations we are using do serve to set upper limits on the squeezing and entanglement available from a more realistic treatment of the physical process~\cite{petard}.

However, given the above caveat, it is instructive to examine the analytical solutions which may be obtained using an undepleted pumps approximation as without a cavity the interaction strengths tend to be small and this approximation is generally very accurate. Setting $\kappa_{1}=\chi_{1}\langle\hat{a}_{4}(0)\rangle$ and  $\kappa_{2}=\chi_{2}\langle\hat{a}_{5}(0)\rangle$ as real positive constants, the Hamiltonian may be written
\begin{equation}
{\cal H}_{int} = i\hbar\left[\kappa_{1}\left(\hat{a}_{1}^{\dag}\hat{a}_{3}^{\dag}-\hat{a}_{1}\hat{a}_{3}\right)+\kappa_{2}\left(\hat{a}_{2}^{\dag}\hat{a}_{3}-\hat{a}_{2}\hat{a}_{3}^{\dag}\right)\right],
\label{eq:undepleteHam}
\end{equation} 
from which we find the Heisenberg equations of motion,
\begin{eqnarray}
\frac{d\hat{a}_{1}}{dt} &=& \kappa_{1}\hat{a}_{3}^{\dag},\nonumber\\
\frac{d\hat{a}_{2}}{dt} &=& \kappa_{2}\hat{a}_{3},\nonumber\\
\frac{d\hat{a}_{3}}{dt} &=& \kappa_{1}\hat{a}_{1}^{\dag}-\kappa_{2}\hat{a}_{2}.
\label{eq:ferraroheisenberg}
\end{eqnarray}
For later convenience we will rewrite the above as equations of motion for the quadrature operators, finding
\begin{eqnarray}
\frac{d\hat{X}_{1}}{dt} &=& \kappa_{1}\hat{X}_{3},\nonumber\\
\frac{d\hat{Y}_{1}}{dt} &=& -\kappa_{1}\hat{Y}_{3},\nonumber\\
\frac{d\hat{X}_{2}}{dt} &=& \kappa_{2}\hat{X}_{3},\nonumber\\
\frac{d\hat{Y}_{2}}{dt} &=& \kappa_{2}\hat{Y}_{3},\nonumber\\
\frac{d\hat{X}_{3}}{dt} &=& \kappa_{1}\hat{X}_{1}-\kappa_{2}\hat{X}_{2},\nonumber\\
\frac{d\hat{Y}_{3}}{dt} &=& -\kappa_{1}\hat{Y}_{1}-\kappa_{2}\hat{Y}_{2}.
\label{eq:Xheis}
\end{eqnarray}
These equations can now be solved analytically to give the solutions for the operators as functions of their initial values, which will all be zero for this system. However, due to bosonic commutation relations, not all the moments vanish  and at $t=0$ with all the output fields as vacuum, we have  $\langle\hat{X}_{i}(0)\hat{X}_{j}(0)\rangle=\langle\hat{Y}_{i}(0)\hat{Y}_{j}(0)\rangle=\delta_{ij}$. 
This is all the information we need to find useful time-dependent solutions for the variances and covariances needed for the correlations which establish tripartite entanglement. 

\subsection{Hyperbolic solutions}

We find that there are three classes of solutions for different regimes, depending on whether $\kappa_{2}^{2}>\kappa_{1}^{2}$, $\kappa_{2}^{2}<\kappa_{1}^{2}$ or $\kappa_{1}^{2}=\kappa_{2}^{2}$. The last of these was treated in Ref.~\cite{nl2} and we will not consider it further here. For $\kappa_{2}^{2}>\kappa_{1}^{2}$, $\Omega$ is imaginary and the solutions are periodic, while for $\kappa_{2}^{2}<\kappa_{1}^{2}$ the solutions are hyperbolic. We will begin with the correlations for the hyperbolic solutions, as this is the operating regime of the Bondani experiment~\cite{Bondani}.

Setting $\Omega=\sqrt{\kappa_{1}^{2}-\kappa_{2}^{2}}$, we find these solutions as
\begin{eqnarray}
\hat{X}_{1}(t) &=& \frac{\kappa_{1}^{2}\cosh\Omega t-\kappa_{2}^{2}}{\Omega^{2}}\hat{X}_{1}(0)-\frac{\kappa_{1}\kappa_{2}\left(\cosh\Omega t-1\right)}{\Omega^{2}}\hat{X}_{2}(0)+\frac{\kappa_{1}\sinh\Omega t}{\Omega}\hat{X}_{3}(0),\nonumber\\
\hat{Y}_{1}(t) &=& \frac{\kappa_{1}^{2}\cosh\Omega t-\kappa_{2}^{2}}{\Omega^{2}}\hat{Y}_{1}(0)+\frac{\kappa_{1}\kappa_{2}\left(\cosh\Omega t-1\right)}{\Omega^{2}}\hat{Y}_{2}(0)-\frac{\kappa_{1}\sinh\Omega t}{\Omega}\hat{Y}_{3}(0),\nonumber\\
\hat{X}_{2}(t) &=& \frac{\kappa_{1}\kappa_{2}\left(\cosh\Omega t-1\right)}{\Omega^{2}}\hat{X}_{1}(0)+\frac{\kappa_{1}^{2}-\kappa_{2}^{2}\cosh\Omega t}{\Omega^{2}}\hat{X}_{2}(0)+\frac{\kappa_{2}\sinh\Omega t}{\Omega}\hat{X}_{3}(0),\nonumber\\
\hat{Y}_{2}(t) &=& -\frac{\kappa_{1}\kappa_{2}\left(\cosh\Omega t-1\right)}{\Omega^{2}}\hat{Y}_{1}(0)+\frac{\kappa_{1}^{2}-\kappa_{2}^{2}\cosh\Omega t}{\Omega^{2}}\hat{Y}_{2}(0)+\frac{\kappa_{2}\sinh\Omega t}{\Omega}\hat{Y}_{3}(0),\nonumber\\
\hat{X}_{3}(t) &=& \frac{\kappa_{1}\sinh\Omega t}{\Omega}\hat{X}_{1}(0)-\frac{\kappa_{2}\sinh\Omega t}{\Omega}\hat{X}_{2}(0)+\hat{X}_{3}(0)\cosh\Omega t,\nonumber\\
\hat{Y}_{3}(t) &=& -\frac{\kappa_{1}\sinh\Omega t}{\Omega}\hat{Y}_{1}(0)-\frac{\kappa_{2}\sinh\Omega t}{\Omega}\hat{Y}_{2}(0)+\hat{Y}_{3}(0)\cosh\Omega t,
\label{eq:XYsols}
\end{eqnarray}
which contain all the information needed to calculate the VLF and OBR correlations in the approximations we are using, except in the case where $\kappa_{1}^{2}=\kappa_{2}^{2}$. In this case the above solutions are not well defined but the equations may still be solved using stochastic integration, as was done in Ref.~\cite{nl2}. For $\kappa_{1}^{2}>\kappa_{2}^{2}$, the time-dependent moments which we need are
\begin{eqnarray}
\langle\hat{X}_{1}^{2}\rangle &=& \langle\hat{Y}_{1}^{2}\rangle = 1+\frac{2\kappa_{1}^{2}}{\Omega^{4}}\left[\kappa_{1}^{2}\sinh^{2}\Omega t+2\kappa_{2}^{2}(1-\cosh\Omega t)\right] \nonumber\\
\langle\hat{X}_{2}^{2}\rangle &=&  \langle\hat{Y}_{2}^{2}\rangle = 1+\frac{1}{\Omega^{4}}\left[2\kappa_{1}^{2}\kappa_{2}^{2}(\cosh\Omega t-1)^{2}\right]\nonumber\\
\langle\hat{X}_{3}^{2}\rangle &=&  \langle\hat{Y}_{3}^{2}\rangle = 1+\frac{2\kappa_{1}^{2}\sinh^{2}\Omega t}{\Omega^{2}}\nonumber\\
\langle\hat{X}_{1}\hat{X}_{2}\rangle &=& -\langle\hat{Y}_{1}\hat{Y}_{2}\rangle = \frac{\kappa_{1}\kappa_{2}}{\Omega^{4}}\left[(\kappa_{1}^{2}+\kappa_{2}^{2})(\cosh\Omega t-1)^{2}+\Omega^{2}\sinh^{2}\Omega t\right]\nonumber\\
\langle\hat{X}_{1}\hat{X}_{3}\rangle &=& - \langle\hat{Y}_{1}\hat{Y}_{3}\rangle = \frac{2\kappa_{1}\sinh\Omega t}{\Omega^{3}}(\kappa_{1}^{2}\cosh\Omega t-\kappa_{2}^{2})\nonumber\\
\langle\hat{X}_{2}\hat{X}_{3}\rangle &=&  \langle\hat{Y}_{2}\hat{Y}_{3}\rangle = \frac{2\kappa_{1}^{2}\kappa_{2}}{\Omega^{3}}(\cosh\Omega t-1)\sinh\Omega t.
\label{eq:moments}
\end{eqnarray}
We note here that the above expectation values are actually the variances and covariances for single modes, as the expectation values of the amplitudes are all zero.

\begin{figure}[tbhp]
\begin{center} 
\includegraphics[width=0.8\columnwidth]{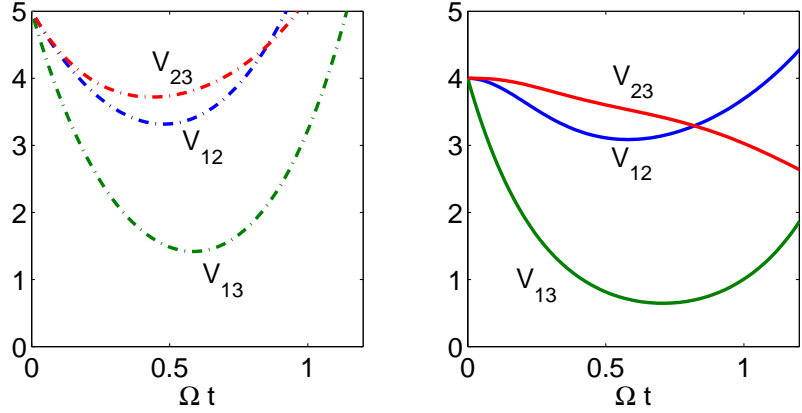}
\end{center} 
\caption{(Colour online) The analytical solutions of the VLF correlations, with $\kappa_{1}=1.2\kappa_{2}$. Any two of the correlations falling below $4$ is sufficient to demonstrate that genuine tripartite entanglement is present. The solid lines use the optimised expressions. All quantities shown in these and subsequent graphs are dimensionless.}
\label{fig:Bondani1}
\end{figure}

\begin{figure}[tbhp]
\begin{center} 
\includegraphics[width=0.8\columnwidth]{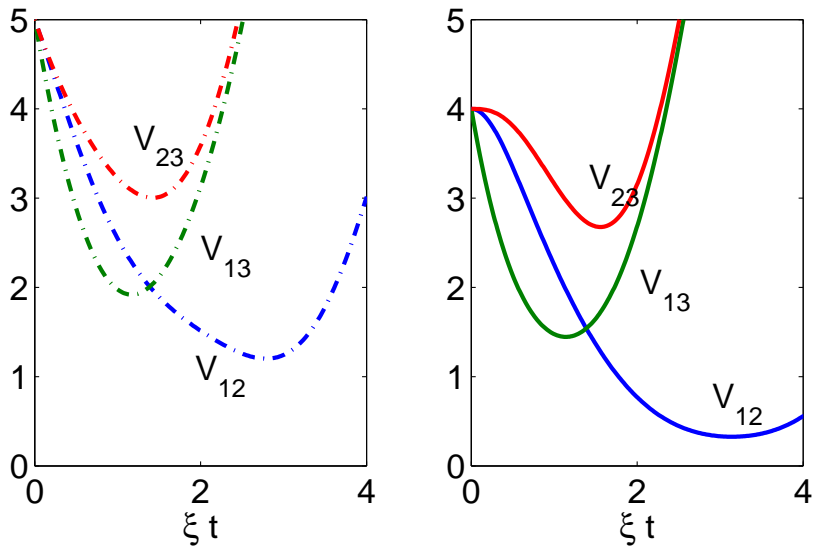}
\end{center} 
\caption{(Colour online) The analytical solutions of the VLF correlations, with $\kappa_{2}=1.8\kappa_{1}$. Any two of the correlations falling below $4$ is sufficient to demonstrate that genuine tripartite entanglement is present. The solid lines use the optimised expressions.}
\label{fig:Bondani2}
\end{figure}

\subsection{Periodic solutions}

We will now look at the case where $\kappa_{2}^{2}>\kappa_{1}^{2}$, which leads to solutions expressed in terms of periodic functions. Setting $\xi=\sqrt{\kappa_{2}^{2}-\kappa_{1}^{2}}$, we find
\begin{eqnarray}
\hat{X}_{1}(t) &=& \frac{\kappa_{2}^{2}-\kappa_{1}^{2}\cos\xi t}{\xi^{2}}\hat{X}_{1}(0)+\frac{\kappa_{1}\kappa_{2}(\cos\xi t-1)}{\xi^{2}}\hat{X}_{2}(0)+\frac{\kappa_{1}\sin\xi t}{\xi}\hat{X}_{3}(0),\nonumber\\
\hat{Y}_{1}(t) &=& \frac{\kappa_{2}^{2}-\kappa_{1}^{2}\cos\xi t}{\xi^{2}}\hat{Y}_{1}(0)-\frac{\kappa_{1}\kappa_{2}(\cos\xi t-1)}{\xi^{2}}\hat{Y}_{2}(0)+\frac{\kappa_{1}\sin\xi t}{\xi}\hat{Y}_{3}(0),\nonumber\\
\hat{X}_{2}(t) &=& \frac{\kappa_{1}\kappa_{2}(1-\cos\xi t)}{\xi^{2}}\hat{X}_{1}(0)+\frac{\kappa_{2}^{2}\cos\xi t-\kappa_{1}^{2}}{\xi^{2}}\hat{X}_{2}(0)+\frac{\kappa_{2}\sin\xi t}{\xi}\hat{X}_{3}(0),\nonumber\\
\hat{Y}_{2}(t) &=& \frac{\kappa_{1}\kappa_{2}(\cos\xi t-1)}{\xi^{2}}\hat{Y}_{1}(0)+\frac{\kappa_{2}^{2}\cos\xi t-\kappa_{1}^{2}}{\xi^{2}}\hat{Y}_{2}(0)+\frac{\kappa_{2}\sin\xi t}{\xi}\hat{Y}_{3}(0),\nonumber\\
\hat{X}_{3}(t) &=& \frac{\kappa_{1}\sin\xi t}{\xi}\hat{X}_{1}(0)-\frac{\kappa_{2}\sin\xi t}{\xi}\hat{X}_{2}(0)+\hat{X}_{3}(0)\cos\xi t,\nonumber\\
\hat{Y}_{3}(t) &=& -\frac{\kappa_{1}\sin\xi t}{\xi}\hat{Y}_{1}(0)-\frac{\kappa_{2}\sin\xi t}{\xi}\hat{Y}_{2}(0)+\hat{Y}_{3}(0)\cos\xi t,
\end{eqnarray}
which lead to the solutions for the moments, 
\begin{eqnarray}
\langle\hat{X}_{1}^{2}\rangle &=& \langle\hat{Y}_{1}^{2}\rangle = 1+\frac{2\kappa_{1}^{2}\left[2\kappa_{2}^{2}(1-\cos\xi t)-\kappa_{1}^{2}\sin^{2}\xi t\right]}{\xi^{4}},\nonumber\\
\langle\hat{X}_{2}^{2}\rangle &=& \langle\hat{Y}_{2}^{2}\rangle = 1+\frac{2\kappa_{1}^{2}\kappa_{2}^{2}(\cos\xi t-1)^{2}}{\xi^{4}},\nonumber\\
\langle\hat{X}_{3}^{2}\rangle &=& \langle\hat{Y}_{3}^{2}\rangle = 1+\frac{2\kappa_{1}^{2}\sin^{2}\xi t}{\xi^{2}},\nonumber\\
\langle\hat{X}_{1}\hat{X}_{2}\rangle &=& -\langle\hat{Y}_{1}\hat{Y}_{2}\rangle = \frac{2\kappa_{1}\kappa_{2}}{\xi^{4}}\left[(\kappa_{1}^{2}+\kappa_{2}^{2})(1-\cos\xi t)-\kappa_{1}^{2}\sin^{2}\xi t\right],\nonumber\\
\langle\hat{X}_{1}\hat{X}_{3}\rangle &=& -\langle\hat{Y}_{1}\hat{Y}_{3}\rangle = \frac{\kappa_{1}}{\xi^{3}}\left[2\kappa_{2}^{2}\sin\xi t-\kappa_{1}^{2}\sin 2\xi t\right],\nonumber\\
\langle\hat{X}_{2}\hat{X}_{3}\rangle &=& \langle\hat{Y}_{2}\hat{Y}_{3}\rangle = \frac{2\kappa_{1}^{2}\kappa_{2}\sin\xi t}{\xi^{3}}\left[1-\cos\xi t\right].
\label{eq:periodics}
\end{eqnarray}

\section{Entanglement results}
\label{sec:resultados}

Analytical expressions can be found for both the VLF and OBR correlations using the results of Eqs.~\ref{eq:moments} and \ref{eq:periodics}, but as these can be rather unwieldy we will present our results graphically. In the interests of compact notation we will use the shorthand $V_{ij}$ for the correlation which contains $V(\hat{X}_{i}-\hat{X}_{j})$.
In Fig.~\ref{fig:Bondani1} we show the results of the VLF correlations for the hyperbolic solutions, with $\kappa_{1}=1.2\kappa_{2}$. The dash-dotted lines are the basic expressions, without any optimisation, and demonstrate that genuine tripartite entanglement is present over a small range of interaction strength. The solid lines (of the same colour online) are the expressions optimised as in Eq.~\ref{eq:opttripart} and are seen to violate the inequalities over a wider range. Perhaps the effect of this optimisation is that it allows for the demonstration of entanglement as soon as the interaction is non-zero, whereas the expressions without optimisation need some finite interaction before any of them go below $4$. This is not a contradiction as entanglement may be present even if the inequalities are not violated, in contrast to the Duan and Simon criteria for Gaussian bipartite systems, which provide necessary and sufficient conditions~\cite{Duan,Simon}. We are not aware of any criteria for tripartite continuous-variable entanglement which provide both necessary and sufficient conditions. In Fig.~\ref{fig:Bondani2} we present results for the same correlations in the regime of periodic solutions, with $\kappa_{2}=1.8\kappa_{1}$. We again see that, as expected, the optimisation procedure allows for demonstration of entanglement over a wider range of interaction strengths.

\begin{figure}[tbhp]
\begin{center} 
\includegraphics[width=0.8\columnwidth]{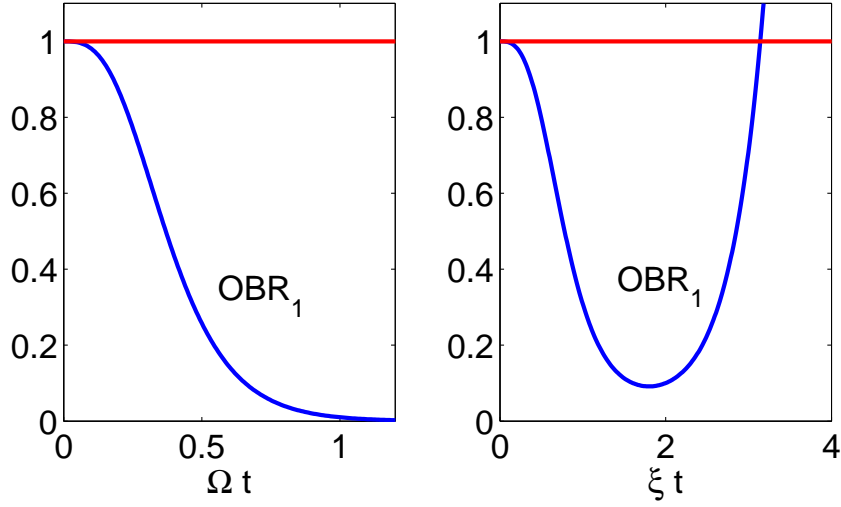}
\end{center} 
\caption{(Colour online) The analytical solutions of the OBR two-mode inference correlations. On the left hand side, $\kappa_{1}=1.2\kappa_{2}$, while on the right hand side $\kappa_{2}=1.8\kappa_{1}$. Although all three correlations should be below one to demonstrate genuine tripartite entanglement, in each case only one of them (OBR$_{1}$) goes below this level, while the other two are both exactly equal to one.}
\label{fig:OBRsingle}
\end{figure}

The three-mode EPR correlations, 
\begin{eqnarray}
OBR_{i} &=& V^{inf}(\hat{X}_{i})V^{inf}(\hat{Y}_{i}),\nonumber\\
OBR_{jk} &=& V^{inf}(\hat{X}_{j}+\hat{X}_{k})V^{inf}(\hat{Y}_{j}+\hat{Y}_{k}),
\end{eqnarray}
may be expressed in terms of the operator moment expectation values using
\begin{eqnarray}
V^{inf}(\hat{X}_{i}) &=& \langle\hat{X}_{i}^{2}\rangle-\frac{[\langle\hat{X}_{i}\hat{X}_{j}\rangle+\langle\hat{X}_{i}\hat{X}_{k}\rangle]^{2}}{\langle\hat{X}_{j}^{2}\rangle+\langle\hat{X}_{k}^{2}\rangle+2\langle\hat{X}_{j}\hat{X}_{k}\rangle},\nonumber\\
V^{inf}(\hat{Y}_{i}) &=& \langle\hat{Y}_{i}^{2}\rangle-\frac{[\langle\hat{Y}_{i}\hat{Y}_{j}\rangle+\langle\hat{Y}_{i}\hat{Y}_{k}\rangle]^{2}}{\langle\hat{Y}_{j}^{2}\rangle+\langle\hat{Y}_{k}^{2}\rangle+2\langle\hat{Y}_{j}\hat{Y}_{k}\rangle},\nonumber\\
V^{inf}(\hat{X}_{j}+\hat{X}_{k}) &=& \langle\hat{X}_{j}^{2}\rangle+\langle\hat{X}_{k}^{2}\rangle+2\langle\hat{X}_{j}\hat{X}_{k}\rangle - \frac{[\langle\hat{X}_{i}\hat{X}_{j}\rangle+\langle\hat{X}_{i}\hat{X}_{k}\rangle]^{2}}{\langle\hat{X}_{i}^{2}\rangle},\nonumber\\
V^{inf}(\hat{Y}_{j}+\hat{Y}_{k}) &=& \langle\hat{Y}_{j}^{2}\rangle+\langle\hat{Y}_{k}^{2}\rangle+2\langle\hat{Y}_{j}\hat{Y}_{k}\rangle - \frac{[\langle\hat{Y}_{i}\hat{Y}_{j}\rangle+\langle\hat{Y}_{i}\hat{Y}_{k}\rangle]^{2}}{\langle\hat{Y}_{i}^{2}\rangle}.
\label{eq:OBRcorrelations}
\end{eqnarray}
In Fig.~\ref{fig:OBRsingle} we give the results for the $OBR_{i}$ in both the periodic and hyperbolic regimes. Neither of these results, which come from inferring the properties of a single quadrature from the properties of a combined two-mode quadrature, gives evidence of genuine tripartite entanglement. In fact, all that these particular correlations succeed in demonstrating is that the combined density matrix, $\rho_{123}$, cannot be separated in the manner $\rho_{123}=\rho_{1}\rho_{23}$, while leaving open the possibilities $\rho_{123}=\rho_{2}\rho_{13}$ and $\rho_{123}=\rho_{3}\rho_{12}$. This shows that choosing to measure these particular criteria to demonstrate entanglement would not be sensible, in contrast to the triply nonlinear system considered in Bradley \etal~\cite{Nosso} and Olsen \etal~\cite{nl3}, where the symmetries of the interaction Hamiltonian meant that any choice of the VLF or OBR criteria was equally useful, with all three giving comparable results.

\begin{figure}[tbhp]
\begin{center} 
\includegraphics[width=0.8\columnwidth]{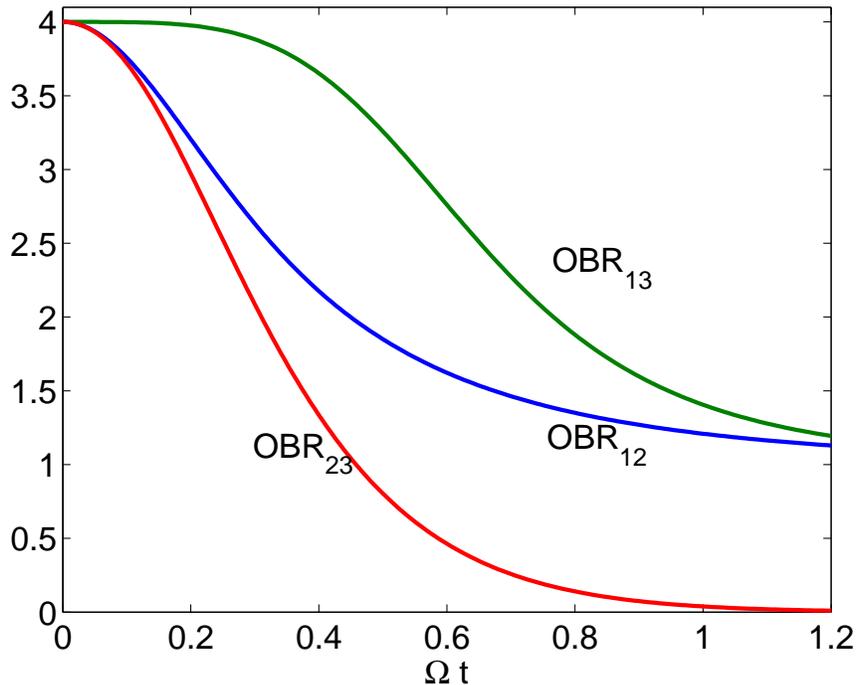}
\end{center} 
\caption{(Colour online) The analytical solutions of the OBR correlations which infer combined mode properties from those of a single mode, with $\kappa_{1}=1.2\kappa_{2}$. All three correlations should be below four to demonstrate genuine tripartite entanglement.}
\label{fig:OBRdouble1}
\end{figure}

\begin{figure}[tbhp]
\begin{center} 
\includegraphics[width=0.8\columnwidth]{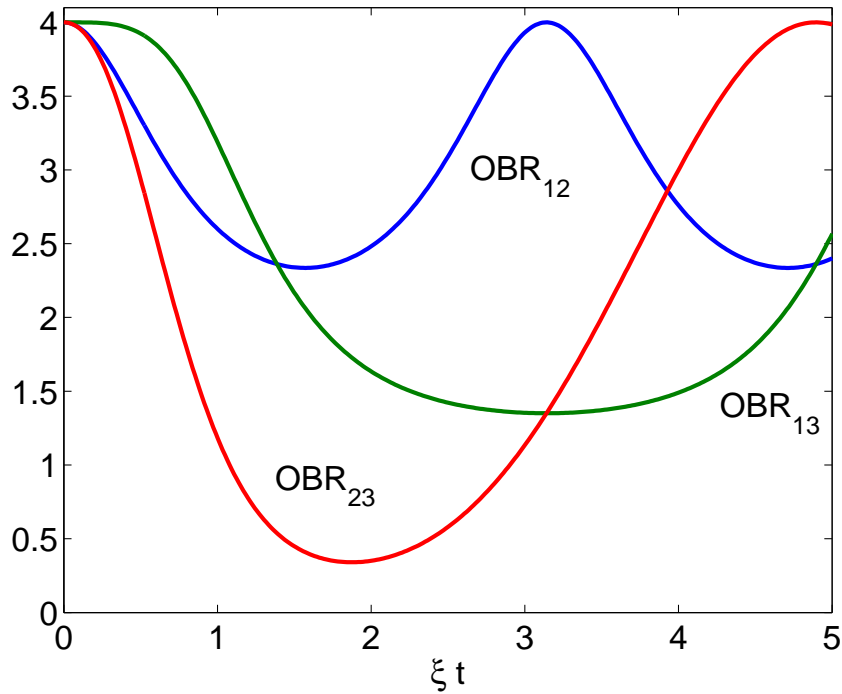}
\end{center} 
\caption{(Colour online) The analytical solutions of the OBR correlations which infer combined mode properties from those of a single mode, with $\kappa_{2}=1.8\kappa_{1}$. All three correlations should be below four to demonstrate genuine tripartite entanglement.}
\label{fig:OBRdouble2}
\end{figure}

However, we do find that with the present system the three-mode EPR correlations ($OBR_{ij}$), which are defined using the properties of one quadrature to infer properties of a combined quadrature which involves the other two modes, are operationally useful. As shown in Fig.~\ref{fig:OBRdouble1} and Fig.~\ref{fig:OBRdouble2}, there is an unambiguous demonstration of the inseparability of the density matrix almost as soon as the interaction begins. This demonstration continues well past the point where the undepleted pumps approximation is expected to lose its validity. Hence, if entanglement were to be demonstrated experimentally with this scheme, measurement of these three correlations would be the preferred option.

\section{Conclusions}

We have examined the interlinked $\chi^{(2)}$ interaction scheme of Bondani \etal~\cite{Bondani} in terms of its suitability for producing output fields which exhibit genuine tripartite entanglement. Using the undepleted pumps approximation, which is valid for small interaction strengths, we have calculated correlations using three different approaches, one which uses the properties of combined quadratures and may be optimised, and the other two which use three-mode generalisations of the EPR argument. We find that these correlations give different answers to the question of whether tripartite entanglement is present in a particular regime and that the most sensitive is that developed by Olsen \etal~\cite{nl3} to infer the properties of a combined mode from those of a single mode. These inequalities are successful in detecting the entanglement over a large regime where the VLF criteria give a false negative. The fact that the correlations give different answers is not contradictory as they all provide sufficient but not necessary criteria and this is a good example of how investigations of continuous variable entanglement become more complicated once we have more than two modes involved.

\begin{acknowledgments}

This research was supported by the Australian Research Council.

\end{acknowledgments}

\end{document}